\documentclass{PoS}

\title{Mega-parsec scale magnetic fields in low density regions in the SKA era:
       filaments connecting galaxy clusters and groups}

\ShortTitle{Mpc scale magnetic fields in low density regions with the SKA}

\author{
Gabriele Giovannini$^{1,2}$, 
Annalisa Bonafede$^3$,
Shea Brown$^{4}$,
Luigina Feretti$^{1}$,
Chiara Ferrari$^{5}$,
Myriam Gitti$^{1,2}$,
Federica Govoni$^{6}$,
Matteo Murgia$^{6}$,
Valentina Vacca$^{7}$
\\ 
$^1$Istituto di Radioastronomia/INAF (BO), I; 
$^2$Dipartimento di Fisica e Astronomia, Bologna University, I;
$^3$Hamburger Sternwarte, Universitat, D;
$^4$Iowa University, USA;
$^5$Lagrange Laboratory, OCA, F;
$^6$Osservatorio Astronomico di Cagliari/INAF, I;
$^7$Max Planck Institute for Astrophysics, D.
\\
E-mail: \email{ggiovann@ira.inaf.it}
}

\abstract{The presence of magnetic fields in galaxy clusters has been
well established in recent years, and their importance for the
understanding of the physical
processes at work in the Intra Cluster Medium has been recognized.
Halo and relic sources have been detected in several tens clusters.
A strong correlation is present between the halo and relic radio power and
the X-ray luminosity. Since cluster X-Ray luminosity and mass are related,
the correlation between the radio power and X-ray luminosity could derive
from a physical
dependence of the radio power on the cluster mass, therefore the cluster mass
could be a crucial parameter in the formation of these sources.
The goal of this  project is to investigate the existence of
non-thermal structures beyond the Mpc scale, and associated with lower density
regions with respect to clusters of galaxies: galaxy filaments connecting
rich clusters. We present a piece of evidence of diffuse radio emission
in intergalactic filaments.
Moreover, we present and discuss the detection of radio emission
 in galaxy groups and in faint X-Ray clusters, to 
analyze non-thermal
properties in low density regions with physical conditions similar to galaxy
filaments.
We discuss how SKA1
observations will allow the investigation of this topic and 
the study of the presence of
diffuse radio sources in low density regions.
This will be a fundamental step to understand the origin and
properties of cosmological magnetic fields.
}

\FullConference{
Advancing Astrophysics with the Square Kilometre Array\\
June 8-13, 2014\\
Giardini Naxos, Sicily, Italy}

\newcommand{\skipthis}[1]{}

\begin{document}

\section{Introduction}

Magnetism is one of the four fundamental forces and plays an important
role in the formation and evolution of objects as large as clusters of
galaxies and as small as stars. In spite of the 
significance of magnetic fields in astrophysics and cosmology, their
origin and properties remain poorly understood.
Observational evidences derived primarily during the last decade
(see e.g. the recent review by Feretti et al. 2012) show
that extended magnetic fields within the Intra Cluster Medium (ICM)
are common in galaxy clusters:\\
-- for $\sim$ 70 clusters deep radio observations have established
the presence of synchrotron extended sources (halos and relics) that are not 
associated
  with individual galaxies, but with the ICM as a whole.  The observed radio
  features imply the existence of $\mu$G magnetic fields on scales 
$\sim$ 1 Mpc;\\
-- Faraday-rotation measures of polarized radio
sources both within and
  behind clusters represent independent probes of the strength of intracluster
  magnetic fields.  Studies on both statistical samples and individual clusters
  (see e.g. Bonafede et al. 2013; Vacca et al. 2012) yield
  consistent values of 1--5 $\mu$G, in cluster central regions.

In addition, evidence has been found for the
existence of magnetic fields on even larger scales and in environments of
lower density, although in many cases the physical nature of the corresponding
structures remains ambiguous.

Here we will present the observational evidence of non-thermal emission in 
galaxy filaments, poor clusters and low mass regions. These
data can improve our knowledge on the origin of cosmological (Mpc scale)
magnetic fields. Finally, we will discuss how important the SKA will be 
for the investigation of this subject.
 
\section{Beyond Galaxy Clusters}

The most convincing evidence of radio emission from a genuine
  filament was presented by Bagchi et al. (2002), 
who found radio emission
  coincident with the filament of galaxies ZwCl 2341.1+0000 (see 
Boschin et al. 2013, for a recent discussion). VLA
  observations of this field (Giovannini et al. 2010) confirmed
  the presence of large-scale diffuse emission extending over an area of
$\sim$3 Mpc in size. The surface brightness in the faint external regions
is $\sim$ 5 $\times$ 10$^{-5}$ mJy/arcsec$^2$ at 1.4 GHz. The structure is 
polarized on average at about 10\% level.

An additional case of diffuse radio emission possibly associated with a
large-scale filament is the Southern component of 0809+39 
(Brown \& Rudnick 2009). 
Its origin is uncertain, though its coincidence with a filament of galaxies 
at z $\sim$ 0.04 suggests that it could be either synchrotron emission from 
filamentary large-scale structure or old emission from an extinct radio 
galaxy.

A complex radio
emission has also been recently found in the A3411-A3412 structure 
(Giovannini et al. 2013, Fig. 1). ROSAT X-ray data show 
a diffuse thermal emission from A3411 and a more compact emission
coincident with A3412. The A3411 emission towards A3412
shows a multiple structure with a few sub-components suggesting 
the presence of an active merger. Optical data confirm the existence
of sub-structures and  
show an extended filament to the SE of A3412, 
aligned with the A3411 -- A3412 system.

The radio image shows a halo source in A3411 and an extended structure in 
between A3411--A3412 and in the region of the SE filament (see Fig. 1
and Giovannini et al. 2013). The
morphology and properties of the diffuse radio emission could be 
explained by electron acceleration and magnetic field amplification originated
from accretion shocks combined with turbulence in the thermal gas due to
the merger between A3411 and A3412 clusters and the SE filament. In particular
the diffuse source SE to A3412, oriented along the giant filament could be 
powered by accretion shocks as material falls 
onto the filament, as it was suggested by Brown \& Rudnick (2011) 
for the relic of the Coma cluster.

\begin{figure}
    \centering
    \includegraphics[width=0.5\textwidth]{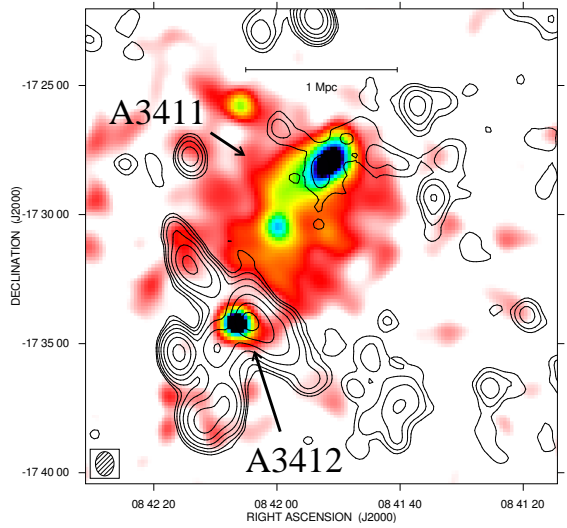}
    \caption{Radio X-ray overlay of the A3411 -- A3412 region. Contours
show the flux density at 1.4 GHz, half power beam width (HPBW) = 
56.3'' x 43.1''. Contours are from 0.12 mJy/beam with a step of
2. In colour the X-ray image from ROSAT data.
Note the extended emission aligned with the A3411--A3412 filament. For more 
details, see Giovannini et al. (2013).
}
   \label{fig:myPlot_1}
\end{figure}

The number of diffuse synchrotron sources connected to galaxy
filaments is presently quite low. This is consistent with the 
relatively high surface brightness limit of present
radio surveys (e.g. NVSS) and with the expected low magnetic-field intensity
in these regions (e.g. 10 nano-Gauss on the volume average for filaments, 
Ryu et al. 2008).
We could speculate that in the few cases where radio emission has been detected
from galaxy filaments, peculiar conditions of merging activity 
among local sub-groups are able to amplify the magnetic fields and accelerate 
relativistic particles (see e.g. Boschin et al. 2013 
for ZwCl 2341.1+0000 and Vazza et al. 2014).

On a smaller scale, 
bridges of diffuse radio emission have been found in a few clusters
connecting radio halos and peripheral relics (Feretti et al. 2012). 
These structures may trace
the presence of synchrotron emission in filaments of merging material into
the main cluster.
The prototype is the bridge
of radio emission connecting the halo Coma-C with the peripheral relic
1253+275 (Kim et al. 1989).
This extended low brightness structure was detected in WSRT data and confirmed
by low resolution single dish observations. 
The surface brightness at 90 cm is lower than $\sim$ 10$^{-4}$ mJy/arcsec$^2$.
Assuming a spectral index of 1.5, a brightness of $\sim$ 2 $\times$ 
10$^{-5}$mJy/arcsec$^2$ at 1.4 GHz is derived.
At present, we do not know in detail its morphology
and properties because of a too low signal to noise ratio
in interferometric data and a too low angular
resolution in single dish images. This structure is visible also in X-ray 
images. Its origin is unknown, but we note that it is in the same direction
of the bridge of  galaxies connecting Coma to A1367 and forming the
super-cluster structure.
A large scale magnetic field is present in the Coma cluster peripheral 
regions (Bonafede et al. 2010, 2013), but the bridge 
properties are quite
different from those of the central halo and of the relic source. 
Turbulence in the ICM could be invoked as the origin of the 
radio emission, but this is problematic because turbulence is not
efficient in accelerating particles in low density regions, and
a connection to shocks is not clear.
Similar features have been found in A2255 and A2744 (Feretti et al. 2012).

\section{Poor clusters and low density regions}

Since cluster X-Ray luminosity and mass are correlated, the correlation between
 radio power and X-ray luminosity in radio halos 
could reflect a dependence of the
radio power on the cluster mass. This correlation could indicate that the
cluster mass is a crucial parameter in the formation of radio halos and
relics.
Since it is likely that massive clusters are the result of several major
mergers, we can conclude that the cluster total mass and past plus present
mergers are the necessary ingredients for the formation and evolution
of diffuse radio sources, in agreement with the result that not
all clusters with recent mergers show a radio halo or a relic source.

This scenario is presently supported by many observational and theroretical
results and it is also in agreement with numerical simulations (see e.g.
Brunetti \& Lazarian 2011, Vazza et al. 2010, Cassano et al. 2010).
However, Brown \& Rudnick (2009, 2011), and Giovannini e al. (2009, 2011)
 have shown the presence of diffuse radio sources in a few clusters with 
low X-ray luminosity, therefore low mass. Since 
low-mass clusters and groups are often found in filaments connecting
rich galaxy clusters and forming a super--cluster structure, it is important 
to increase our knowledge of these sources.

The best known case is the poor group of galaxies 0217+70, where an
extended central radio halo and double peripheral relic radio
sources are present (Brown et al. 2011).
Further cases are the radio halo in A523 and in A1213 (Giovannini et al. 2009).

We present in Fig. 2-Left  the plot of the total radio power 
and X-ray luminosity 
for radio halos as shown in Feretti et al. (2012), where X-ray underluminous 
clusters are indicated by their name. 
These halos are 
overluminous in radio by at least an order of magnitude with respect to the
expectation  of the  radio power -- X-ray 
luminosity correlation.
This new population of diffuse radio halos
emission opens up the possibility of probing the correlation between
low-mass cluster mergers and non-thermal cluster properties (particle
acceleration and diffuse magnetic fields) with
upcoming deep radio continuum surveys.

\begin{figure}
    \centering
    \includegraphics[width=0.45\textwidth]{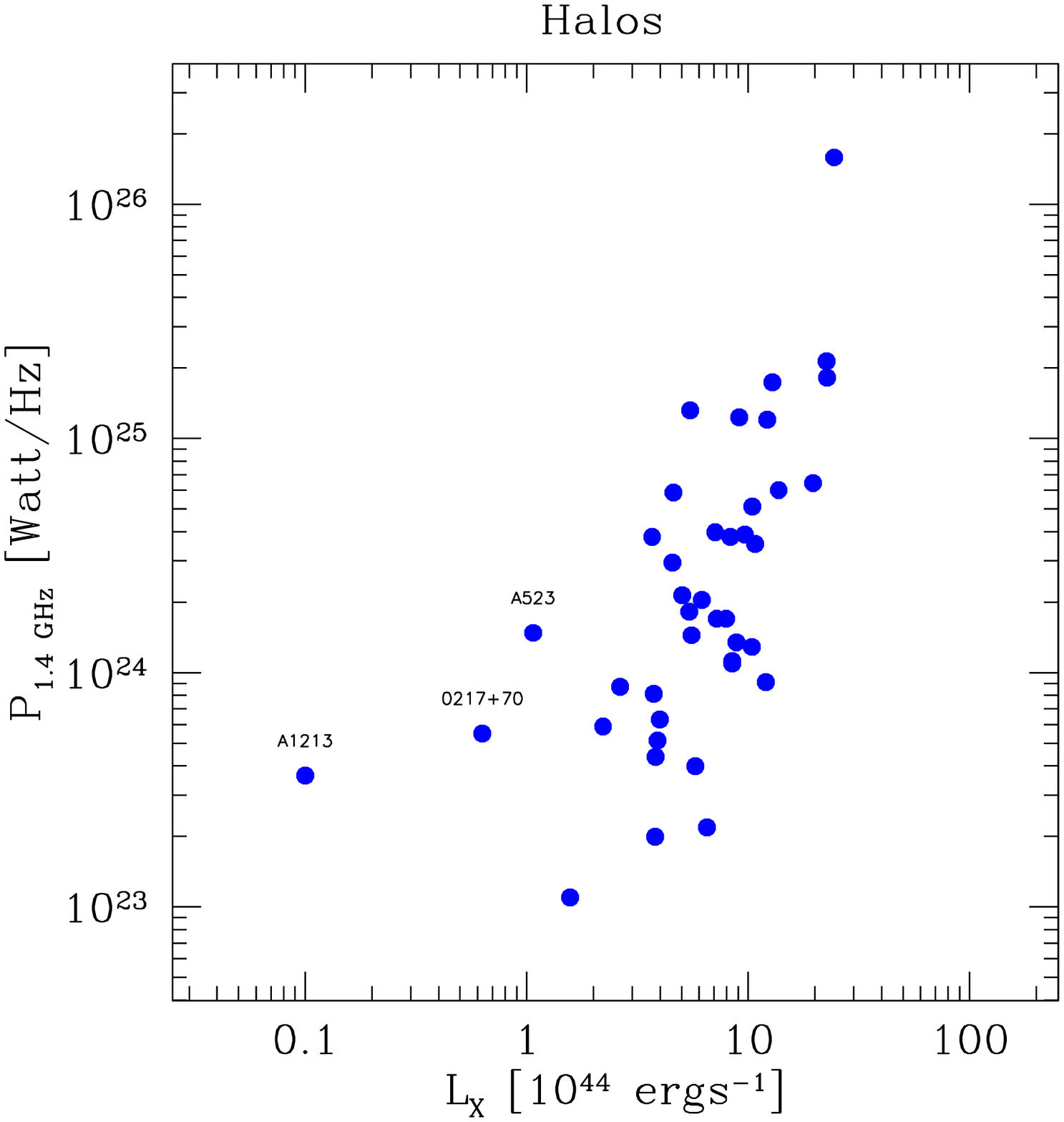}
    \includegraphics[width=0.45\textwidth]{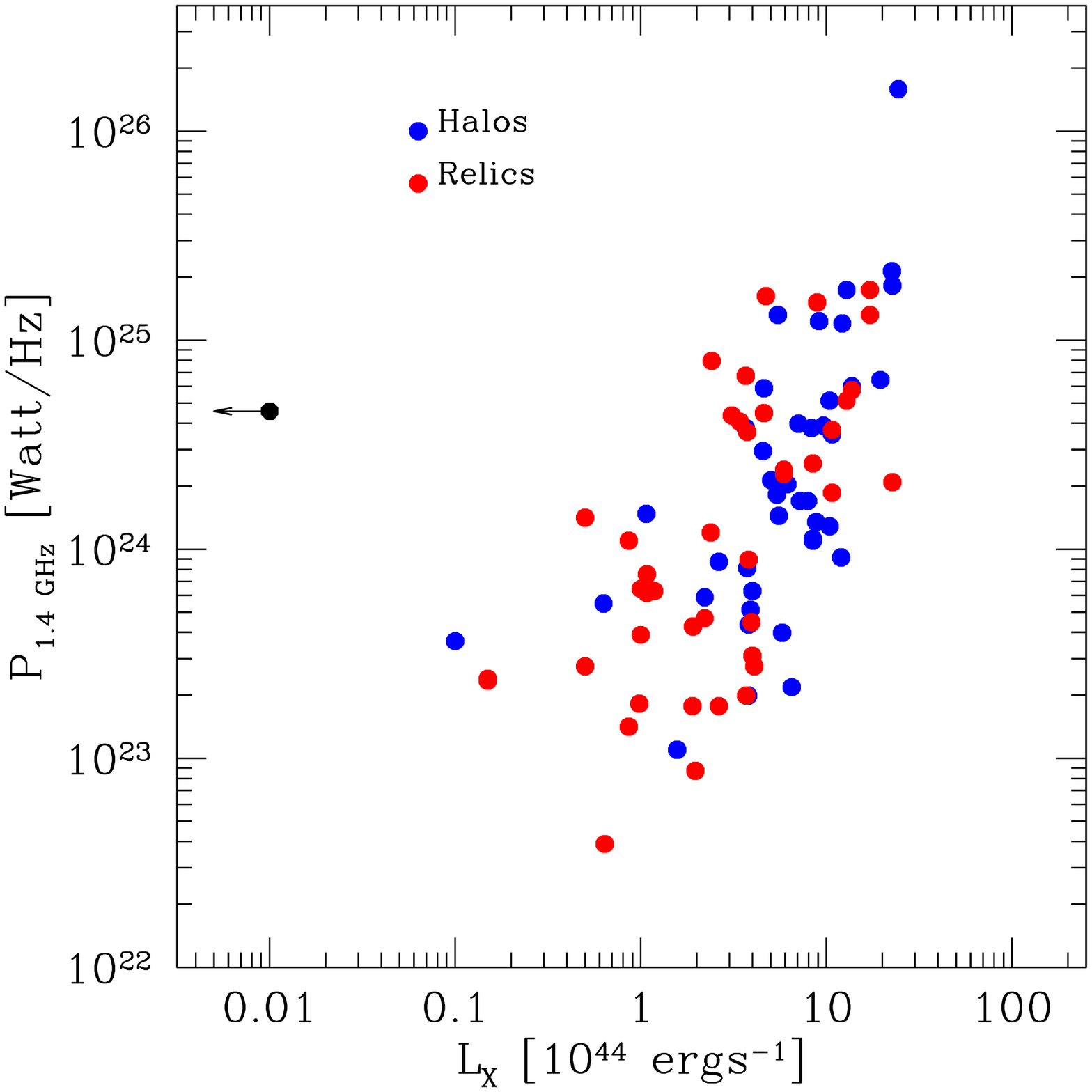}
    \caption{Left: monochromatic radio power of radio halos at 1.4 GHz versus 
the cluster X-ray luminosity between 0.1 and 2.5 keV for merging clusters with
radio halos. Underluminous X-ray clusters are named (from Feretti et al. 2012).
Right: monochromatic radio power of radio halos (blue) and relics (red) at 
1.4 GHz versus
the cluster X-ray luminosity between 0.1 and 2.5 keV for merging clusters
(data are from Feretti et al. 2012).
The peculiar source 0917+75 is shown in black.}
   \label{fig:myPlot_2}
\end{figure}

\section{Two peculiar structures: 0917+75 and A399-A401}

The radio source 0917+75 (Fig. 3) is an elongated diffuse emission
studied by Dewdney et al.\ (1991); Harris et al.\ (1993), and Giovannini \&
Feretti (2000).
It is located in a region
away from rich clusters, the nearest one being A786 (z = 0.124) at $\sim$ 4 
Mpc.
This distance is approximately twice the A786 virial radius, too
large for the radio
  emission being the result of present or past interaction with the cluster.
In this region, a few galaxy clusters at the same redshift are present 
(A787, A762, A748, A765), which  belong to the Rood
Group of clusters of galaxies N. 27. Their
very large distance from the source favours the hypothesis that the diffuse
radio structure is related to a galaxy group
or to the supercluster structure.
Dewdney et al. (1991) noted that a group of galaxies
with an apparent magnitude similar to that of galaxies in
A786 is present in the
region of the diffuse radio emission and obtained the redshift of the two
brightest galaxies (0.1241 and 0.1259) confirming the association.
\begin{figure}
    \centering
    \includegraphics[width=0.5\textwidth]{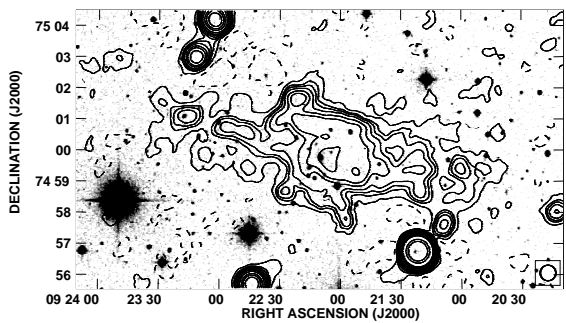}
    \caption{Isocontour image of the diffuse source 0917+75 at 1.4 GHz and 
HPBW = 20''. Contour levels are: -0.2 0.2 0.5 0.7 1 1.5 2 3 5 7 10 30 50 
mJy/beam, overimposed on the optical image from the DPSS.
}
   \label{fig:myPlot_3}
\end{figure}
Girardi et al. (in progress) studied the distribution of
galaxies of the surrounding large scale structure using the 
SuperCOSMOS Sky Survey. The distribution of
the photometric members of the supercluster shows the close
clusters as clear overdensities, while several 
faint galaxies are present in the 0917+75 region, around the two
brightest galaxies observed by Dewdney et al. (1991), 
confirming the presence of a group or a filament. 
The total spectral index is complex as discussed by Harris et al. (1993), 
showing an
average value between 150 MHz and 1.5 GHz of $\sim$ 1.5, with a possible high
frequency steepening.
The source is polarized at a 25\% level at 1.4 GHz in the brightest inner
region (Giovannini et al., in preparation), unexpectedly high for a radio halo.

The main peculiarity of this source is 
the lack of  X-ray emission (see Fig. 2-Right). This  is unexpected 
in comparison
with other Mpc scale diffuse cluster radio sources (halos or relics), where
the radio power is correlated to the X-ray luminosity.
X-ray upper limits obtained with the XMM--Newton observatory, 
impose upper limits on the X-ray flux due to inverse Compton scattering
of photons from the cosmic microwave background by relativistic electrons in 
the diffuse source and imply that the
local magnetic field has to be $>$ 0.81 $\mu$Gauss (three sigma level,
Chen et al. 2008).
This makes this source quite unique.
We suggest that this source could be an inherent feature of the filamentary
structure of the Rood 27 supercluster, even if the measured magnetic field
lower limit is higher than expected in filamentary structures according
to numerical simulations (see e.g. Ryu et al. 2008). Alternatively it could 
be the remnant of a died radio galaxy. If confirmed it will suggest that
AGN activity is important to spread seeds of relativistic particles and 
magnetic fields in filaments on a scale as large as $\sim$ 2 Mpc.

A399 and A401 is a  pair of galaxy clusters separated in 
projection by an angular
distance of 36', corresponding to a linear separation of 3 Mpc. 
It represents the first example of a double radio halo in the close 
pair of galaxy
clusters (Murgia et al. 2010, Fig. 4). The discovery of this
double halo is exceptional, owing to the rarity of halo radio sources in 
general, and given that
X-ray data (e.g. Sakelliou \& Ponman 2004, Fujita et 
al. 2008),
seem to suggest that the two clusters are still in a pre-merger state.
The existence of these radio halos 
leads naturally to the question as to whether faint radio emission
related to weaker magnetic fields could be present 
in the intergalactic medium between the two clusters, namely on the
scale of galaxy filaments. This question is supported by 
new Planck satellite data that revealed 
a Sunyaev--Zeldovich (SZ)
signal between A399 and A401 (Planck 
Collaboration, 2013). The intercluster
SZ signal is compatible with a scenario where the intercluster region is 
populated with a mixture
of material from the clusters and the intergalactic medium, indicating the
presence of a 
bridge of matter connecting the two systems. 
It may be an evidence that the region is witnessing the process of a 
large--scale structure formation, where cosmic shocks originated by complex 
merger events are able
to amplify magnetic fields and accelerate synchrotron electrons.

\begin{figure}
    \centering
    \includegraphics[width=0.8\textwidth]{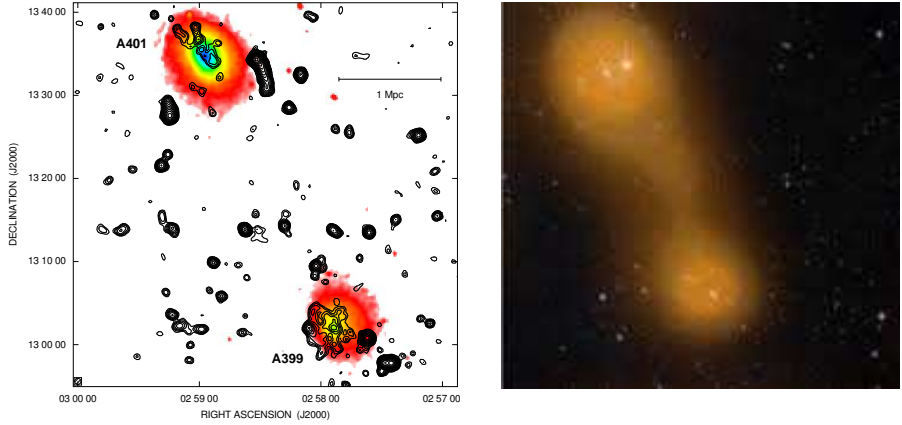}
    \caption{Left: Total intensity radio contours at 1.4 GHz of the system 
A399--A401 (Murgia et al. 2010), overlaid on the XMM X-ray image. Right: 
Sunyaev-Zeldovich emission (Planck collaboration 2013) showing a bridge of 
hot gas connecting A399--A401.
}
   \label{fig:myPlot_4}
\end{figure}

\section{The Future}

To improve our knowledge on the origin and properties of Mpc scale magnetic 
fields we need a comprehensive view of the radio emission
on very large scales, to complement the information obtained from
 the well known  halos and 
relics present  in high X-ray luminous merging clusters: 

1) non-thermal emission in larger structures, going from radio halos 
in clusters to the
radio bridges connecting clusters to galaxy filaments and superclusters;

2) non-thermal emission from galaxy clusters with low mass and low X-ray 
luminosity, often present in supercluster structures connecting rich
clusters.

From the surface brightness in the few galaxy filaments 
discussed above, we can estimate the needs of the 
expected new observations
with SKA1 for a deeper understanding of large scale magnetic fields.
The two prototypes of these sources are considered to be
ZwCl2341.1+0000 and the bridge in
the Coma cluster.
In ZwCl2341.1+0000, the surface brightness in fainter diffuse
regions is of the order of 0.05 $\mu$Jy/arcsec$^2$ at 1.4 GHz.
The bridge of radio emission connecting the Coma-C radio halo with the Coma
relic source 1253+275 has been detected only at 327 MHz (Giovannini et al.
1990) and its
surface brightness $\sim$ 0.02 $\mu$Jy/arcsec$^2$ at 1.4 GHz,
assuming a spectral index of 1.5 between 327 and 1415 MHz.
From these data we estimate that a sensitivity of $\sim$ 
10$^{-2}$ $\mu$Jy/arcsec$^2$ at 1.4 GHz are needed. 
In addition, high angular resolution is necessary to avoid confusion 
limited images.
The required compromise between high surface brightness sensitivity and high 
resolution is best achieved with SKA1-MID (with respect to SKA1-SUR).
Comparing Key System Performance Factors for SKA1-MID Array (Table 6 and 7 in 
Dewdney et al. 2013), 
with the confusion level estimated by Condon et al. (2012),
the best choice would be to observe in Band 2 with the maximum available
bandwidth (0.8 GHz) and an angular resolution of 
$3''$. The minimum detectable flux density in a 2 hour observation should be
$\sim$ 0.08 $\mu$Jy/beam, at the same level of the noise-free confusion limit 
= 0.08 $\mu$Jy at this angular resolution. Since the regions with the lowest
surface brightness (assuming the conservative hypothesis
of uniform distribution) in our sources are at a level of 0.5 -- 0.2 
$\mu$Jy/beam, 
suggested observations should be able to detect likely at 2.5--6 sigma level 
these 
kind of radio structures in all covered sky.

To investigate the chance to detect faint halos, we used the 
synthetic radio halos generated by Xu et al. (2012), under
the assumption that the 
initial magnetic fields are injected into the ICM  by AGN at high redshift.
Assuming energy
equipartition between magnetic fields and non-thermal electrons, Xu et al. 
compared the global properties of the mock radio
halos with the real ones (see Fig. 5). 
The synthetic radio halos agree with the observed correlations 
between the radio power versus the cluster
X-ray luminosity and between the radio power versus the radio
halo size. Out of 16 simulated
clusters, eight present a radio halo detectable by present 
radio telescopes (e.g. JVLA) under the hypothesis of equipartition condition. 
The eight simulated clusters whose
radio emission is below the detection threshold will need deeper
radio observations. Using upper limit on
total radio power P$_{1.4 GHz}$ for the undetected radio halos, 
assuming a putative size as derived 
from the total radio power versus the largest linear size
 correlation, and an average surface brightness of the diffuse
emission uniform over the putative halo size,
the undetected clusters should host at z = 0.2 radio halos with a 
1.4 GHz surface
brightness $\sim$ 0.3 $\mu$Jy/beam, assuming a 3'' HPBW.
Therefore, also in this case, SKA1-MID observations in 2 hours 
should be able to
detect at 4 sigma level these faint radio halos up to at least z = 0.2 
(see Fig. 5).

\begin{figure}
    \centering
    \includegraphics[width=0.8\textwidth]{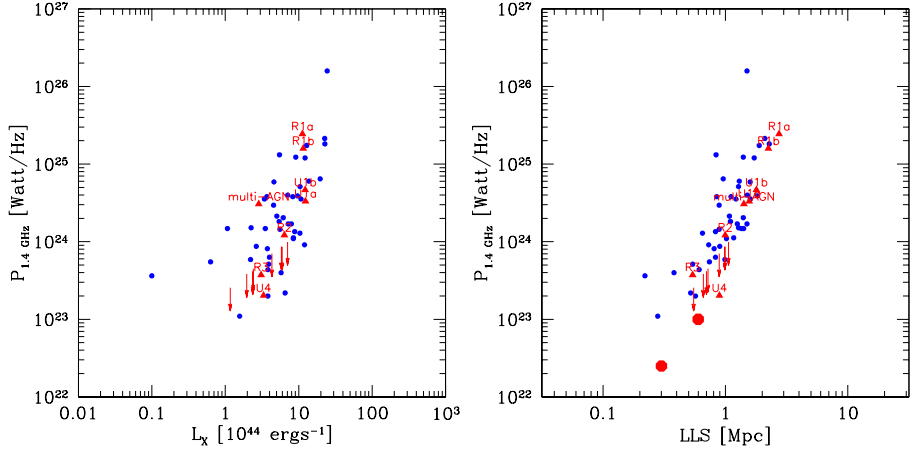}
    \caption{Left: Radio power of radio halos at 1.4 GHz versus the cluster 
X-ray luminosity in the 0.1--2.4 keV band. Right: radio power of 
halos at 1.4 GHz
versus their largest angular size (LLS). Blue dots are observed clusters,
red triangles are simulated clusters, while arrows indicate upper limits
on the radio power of simulated clusters (from Xu et al. 2013). Red circles
on the Right panel are radio halos detectable at 4 sigma level with SKA1-MID 
Band 2 at
z = 0.2, as discussed in the text.
}
   \label{fig:myPlot_4}
\end{figure}

A caveat is that  the angular size of radio
emission connected to galaxy filaments is unknown. 
Moreover, low power sources are
best detected at low redshift because of higher flux density, but they
could show a large angular size.
We note that the bridge of emission in the Coma cluster, as well 
as the Coma-C radio halo,
are not detected by JVLA at 1.4 GHz, not because of sensitivity limit,
but because of missing short spacings (the JVLA shortest baseline 
is $\sim$ 40 meters).
To study this kind of structures at this frequency, SKA1-MID should have
baselines shorter than 20 m.
Observations at low frequencies (SKA1-MID in Band-1, e.g. 700 MHz, and 
SKA1-LOW) are very suitable to detect this diffuse, steep spectrum radio
emission. However, a resolution of a few arcsec is required to avoid confusion
limit problems. With a maximun baseline of 100 km, a HPBW $\le$ 5'' is reached
down to 150 MHz, thus allowing these observations. The SKA1-LOW requested 
sensitivity is $\sim$ 10 $\mu$Jy/beam.

At higher frequencies (i.e. SKA1-MID, Band 3 or 4), these sources are not 
detected because of missing short spacings, therefore a combination with 
single-dish data is necessary. 

In conclusion, with these observing specifications we expect to be able to 
estimate the magnetic 
field strength and radial profile both from synchrotron emission and
 from Faraday Rotation measure
of diffuse and discrete sources within or behind clusters.

Moreover, the large bandwidth will allow to study with a single observation
the spectral index 
distribution of diffuse sources to derive the origin and properties of 
relativistic particles and acceleration mechanism. Comparison with X-ray 
images will permit to individuate reacceleration
regions and the presence of shocks in the intra-cluster medium. 




\vskip 0.5truecm
\vfill\eject
\noindent
{\bf References}
\vskip 0.5truecm

\noindent
Bagchi, J., En{\ss}lin, T.~A., Miniati, F., et al.,\ 2002, New Astronomy, 7,
249 

\noindent
Bonafede, A., Feretti, L., Murgia,
M., et al.: A\&A 513, 30

\noindent Bonafede, A., Vazza, F., 
Br{\"u}ggen, M.,  et al.: 2013 MNRAS 433, 3208

\noindent
Boschin, W., Girardi, M., Barrena, 
R.: 2013 MNRAS 434, 772

\noindent
 Brown, S. \& Rudnick, L.: 2009 
AJ 137, 3158

\noindent 
Brown, S. \& Rudnick, L.: 2011 
MNRAS 412, 2 

\noindent
Brown, S., Duesterhoeft, J., Rudnick, 2011, ApJ 727, L25

\noindent
Brunetti, G. \& Lazarian, A.: 2011 MNRAS 412, 817 

\noindent
Cassano, R., Ettori, S., Giacintucci, S., et al.,: 2010 ApJ,
721, 82

\noindent 
Chen, C.M.H., Harris, D.E., Harrison,
F.A., et al., 2008 MNRAS 383, 1259

\noindent 
Condon, J.J., Cotton, W.D., Fomalont,
E.B., et al. 2012 ApJ 758, 23

\noindent 
Dewdney, P.E., Costain, C.H., 
McHardy, I., et al. 1991 ApJS 76, 1055

\noindent 
Dewdney, P.E., Turner, W., Millenaar, R., et al.
2013 SKA1 System Baseline Design SKA-TEL-SKO-DD-001

\noindent
Feretti, L., Giovannini, G., 
Govoni, F., Murgia, M.: 2012, Astron. \& Astrophys. Rev. 20, 54 

\noindent  
Fujita, Y., Tawa, N., Hayashida, K.,
et al.: 2008 PASJ 60, 343

\noindent
Giovannini, G., \& Feretti, L.: 2000 New Astronomy, 5, 335 

\noindent
Giovannini, G., Kim, K. T., Kronberg, P. P., Venturi, T.: 1990 IAUS 140,
Edrs. R. Beck, P.P. 

Kronberg, R. Wielebinski p.492

\noindent
Giovannini, G., Bonafede, A., Feretti, L., et al.: 2009 A\&A, 507, 1257

\noindent
Giovannini, G., Bonafede, A., Feretti, L., et al.,\ 2010 A\&A, 511, L5 

\noindent
Giovannini, G., Feretti, L., Girardi, M., et al.: 2011 A\&A, 530, 5

\noindent
Giovannini, G., Vacca, V., Girardi, M., 
Feretti, L., et al. 2013 MNRAS 435, 518

\noindent
Harris, D.~E., Stern, C.~P., Willis, A.~G., et al.: 1993 AJ, 105,
769 

\noindent
Kim, K.-T., Kronberg, P.~P., Giovannini, G., et al.,\ Nature, 341,
720 (1989)

\noindent
Murgia, M., Govoni, F., Feretti, L., Giovannini, G.: 2010 A\&A 509, 86

\noindent 
Ryu, D., Kang, H., Cho, J., Das, S.: 2008 Science 320, 909

\noindent 
Sakelliou, I., \& Ponman, T. 2004
MNRAS 351, 1439

\noindent 
Planck collaboration VIII,
2013 A\&A 550, 134

\noindent 
Vacca, V., Murgia, M., Govoni, F., Feretti, L., et al. 2012 A\&A 540, 38

\noindent Vazza F., Brunetti, G., Gheller, C.,
Brunino, R.: 2010 NewA 15, 695

\noindent
Vazza F., Ferrari C., Bonafede A., et al.: 2014 ``Filaments of the radio
cosmic web: opportunities 

and challenges for SKA'', in proceedings of 
``Advancing Astrophysics with the Square 

Kilometer Array''. PoS(AASKA14)097

\noindent 
Xu, H., Govoni, F., Murgia, M., et al.:
2012 ApJ 759, 40

\end{document}